\documentclass{PoS}

\newcommand{\beq}{\begin{eqnarray}}
\newcommand{\eeq}{\end{eqnarray}}

\title{Search for the IR fixed point in the twisted Polyakov loop scheme}

\ShortTitle{Search for the IR fixed point in the twisted Polyakov loop scheme}

\author{Erek Bilgici$^a$, Antonino Flachi$^b$, \speaker{Etsuko Itou}$^c$,Masafumi Kurachi$^{d}$, C.-J. David Lin$^e$, Hideo Matsufuru$^f$, Hiroshi Ohki$^{b,g}$, Tetsuya Onogi$^h$, Eigo Shintani$^h$ and Takeshi~Yamazaki$^i$\\
$(a)$Institut f\"{u}r Physik, Universit\"{a}t Graz, A-8010 Graz, Austria\\
$(b)$Yukawa Institute for Theoretical Physics, Kyoto University, Kyoto 606-8502,
Japan\\
$(c)$Academic Support Center, Kogakuin University, Nakanomachi Hachioji, 192-0015,
Japan\\
$(d)$Theoretical Division T-2, Los Alamos National laboratory, Los Alamos, NM 87544,
USA\\
$(e)$National Chiao-Tung University, and National Center for Theoretical Sciences, Hsinchu 300, Taiwan\\
$(f)$High Energy Accelerator Research Organization (KEK), Tsukuba 305-0801, Japan\\
$(g)$Department of Physics, Kyoto University, Kyoto 606-8501, Japan\\
$(h)$Department of Physics, Osaka University, Toyonaka 560-0043, Japan \\
$(i)$Center for Computational Sciences, 
University of Tsukuba, Tsukuba, Ibaraki 305-8577, Japan\\
E-mail: \email{itou@yukawa.kyoto-u.ac.jp}
}

\abstract{
We present a non-perturbative study of the running
coupling constant in the Twisted Polyakov Loop (TPL) scheme.
We investigate how the systematic and statistical errors can
be controlled {\it via} a feasibility study in SU(3) pure Yang-Mills
theory.
We show that our method reproduces the perturbative
determination of the running coupling in the UV.  In addition, our numerical
result agrees with the theoretical prediction of this coupling
constant in the IR.
We also present our preliminary results for $N_f=12$ QCD, where an IR fixed point may be present.
}

\FullConference{The XXVII International Symposium on Lattice Field Theory - LAT2009\\
		 July 26-31 2009\\
		 Peking University, Beijing, China}

\begin{document}

\section{Introduction}

The existence of non-trivial fixed points is one of most intriguing subjects in quantum field theory.
Field theories with fixed points exhibit scale-invariant behaviour and are often exactly solvable. 
In addition, renormalization group (RG) flows around fixed points contain information of the universality class 
of field theories.

Non-trivial fixed points have been identified non-perturbatively in two-dimensional scalar field theories,
{\it via} the techniques of algebraic method or Wilsonian RG. In  the case of three-dimensional scalar
field theories, they can be found using the large-N expansion or Wilsonian RG.
In four dimensions, there is no non-trivial fixed point for scalar field theories.
On the other hand, for gauge theories in four dimensions, there are Gaussian fixed points.
Furthermore, the perturbative $\beta$ function indicates the existence of non-trivial infrared (IR) 
fixed points for a certain region of large-flavor ($N_f$) SU(N) gauge theories.
Possible appearance of these IR fixed points has stimulated phenomenological studies of topics such as
dynamical electro-weak symmetry breaking and unparticle physics.
The existence of these IR fixed points depends on the gauge group, the number of flavours, and the representation 
of fermion fields.
For SU(3) gauge theory with fermions in the fundamental representation, such a fixed point has been 
predicted in the range $8< N_f \le 16$ using 
perturbation theory \cite{Banks:1981nn}.
However, the value of the renormalized coupling at the point depends on $N_f$, and it may be in
the regime where perturbation theory is not applicable.
Therefore it is important to investigate the existence of this IR fixed point non-perturbatively.

First such lattice study for SU(3) gauge theory was carried out in Ref.~\cite{Damgaard:1997ut}, where the authors investigated 
the phase structure of the case of $N_f=16$.
Recently, Appelquist {\it et al.} performed lattice calculation of the running coupling constant in the 
Schr\"{o}dinger functional (SF) scheme and discovered evidence of an IR fixed point in the case of 
$N_{f} = 12$~\cite{Appelquist:2009ty}.
In their work, no such evidence was found for $N_f=8$.
Furthermore, two groups have studied the phase structure of the $N_f=12$ theory \cite{Deuzeman:2009mh,Fodor:2008hn}.
However, the existence of IR fixed point is not firmly established yet~\cite{Fodor:2008hn}.
The difficulty is mainly due to scheme dependence of the running coupling constant and the presence of significant
lattice artifacts in the strong-coupling regime.
Therefore it is important to measure the running coupling in different renormalisation schemes.

In this work, we perform lattice simulation of the running coupling constant for the fundamental-representation,
$N_f = 12$, SU(3) gauge theory.
Similar to the approuch of Appelquist {\it et al.}, we measure the step scaling function $\sigma (s, g^2(L))=g^2(sL)$ 
keeping the values of $\beta$ that give constant $g^2(L)$ for each small lattice size.
We work in the Twisted Polyakov Loop (TPL) scheme which does not contain $O(a/L)$ discretization errors.
These errors are present in the SF scheme due to the boundary counterterm.
This TPL scheme was first proposed by de Divitiis {\it et al.} \cite{deDivitiis:1993hj,deDivitiis:1994yp} for SU(2) gauge theory,
and we extend the definition of the scheme to the SU(3) case.

In this paper, we give a short review of TPL scheme in \S.\ref{sec:TPL}. 
In \S.\ref{sec:quenched} we present a validity study of this scheme by calculating the running coupling 
constant in SU(3) pure Yang-Mills theory.
Our preliminary results for $N_f=12$ SU(3) gauge theory is reported in \S.\ref{sec:Nf-12}.

\section{Twisted Polyakov Loop scheme}\label{sec:TPL}
In this section, we present the definition of the Twisted Polykov Loop scheme in SU(3) gauge theory.
This is an extension of the SU(2) case as discussed in Ref.~\cite{deDivitiis:1993hj}.
To define the TPL scheme, we introduce twisted boundary condition for the link variables in $x$ and $y$ directions on the lattice:
\beq
U_{\mu}(x+\hat{\nu}L/a)=\Omega_{\nu} U_{\mu}(x) \Omega^{\dag}_{\nu}. \hspace{0.5cm}  (\nu=1,2) \label{twisted-bc-gauge}
\eeq
Here, $\Omega_{\nu}$ are the twist matrices which have the following properties:
\beq
\Omega_{1}\Omega_{2}=e^{i2\pi/3}\Omega_{2}\Omega_{1}, \Omega_{\mu} \Omega_{\mu}^{\dag}=1, (\Omega_{\mu})^3=1, \mbox{Tr}[\Omega_{\mu}]=0.
\eeq
The gauge transformation $U_\mu (r) \rightarrow \Lambda (r) U_\mu (r) \Lambda^\dag (r+\hat{\mu})$ and eq.(\ref{twisted-bc-gauge}) imply 
\beq
\Lambda (r+ \hat{\nu}L/a)=\Omega_\nu \Lambda(r) \Omega_\nu^\dag.
\eeq
Because of this twisted boundary condition, the definition of Polyakov loops in the twisted directions are modified,
\beq
P_{1}(y,z,t) ={\mbox{Tr}} \left( [ \prod_{j} U_{1}(x=j,y,z,t)] \Omega_{1} e^{i2\pi y/3L} \right),
\eeq
in order to satisfy gauge invariance and translation invariance.
The renormalized coupling in TPL scheme is defined by taking 
the ratio of Polykov loop correlators in the twisted ($P_1$) and the untwisted ($P_3$) directions:
\beq
g^2_{TP}=\frac{1}{k} \frac{\langle \sum_{y,z} P_{1} (y,z,L/2a) P_{1} (0,0,0)^{\dag} \rangle}{ \langle \sum_{x,y} P_{3} (x,y,L/2a) P_{3} (0,0,0)^{\dag} \rangle }.\label{TPL-def}
\eeq 
At tree level, this ratio Polyakov loops is proportional to the bare coupling.  The
proportionality factor $k$ is obtained by analytically calculating the one-gluon-exchange diagram.
To perform this analytic calculation, we choose the explicit form of the twist matrices~\cite{Trottier:2001vj},
\beq
\Omega_1=\left( 
\begin{array}{ccc}
0 & 1 & 0\\
0 & 0 & 1\\
1 & 0 & 0
\end{array}
\right),
\Omega_2=\left( 
\begin{array}{ccc}
e^{-i2\pi /3} & 0 & 0\\
0 & e^{i2\pi /3} & 0\\
0 & 0 & 1
\end{array}
\right).
\eeq 
In the case of SU(3) gauge group,
\beq
k&=&\frac{1}{24 \pi^2} \sum \frac{(-1)^n}{n^2+(1/3)^2}\nonumber\\
&=&\frac{1}{24 \pi^2} \left[ \frac{9}{2}-\frac{3\pi}{2} cosech \left(\frac{\pi}{3} \right) \right]\nonumber\\
&=&0.03184\cdots .
\eeq

The naive twisted boundary condition for lattice fermions can be written by
\beq
\psi (x+\hat{\nu}L/a)=\Omega_{\nu} \psi(x).
\eeq
However, this results in an inconsistency when changing the order of translations, namely,
\beq
\psi (x+\hat{\nu}L/a+\hat{\rho}L/a)&=&\Omega_{\rho} \Omega_{\nu} \psi(x), \nonumber\\
&\ne &\Omega_{\nu} \Omega_{\rho} \psi(x).
\eeq  
To avoid this difficulty, we introduce "smell" symmetry~\cite{Parisi:1984cy}, which is a copy of color symmetry.
We indentify the fermion field as a $N_c \times N_s$ matrix ($\psi^a_\alpha$(x)).
Then we impose the twisted boundary condition for fermion fields to be 
\beq
\psi^a_{\alpha} (x+\hat{\nu}L/a)= e^{i \pi/3} \Omega_{\nu}^{ab} \psi^{b}_{\beta} (\Omega_{\nu})^\dag_{\beta \alpha}
\eeq
for $\nu=1,2$ directions.
Here, the smell index can be considered as a ``flavor'' index, then the number of flavors should be a multiple of $N_s(=N_c=3)$.
We use staggered fermion in our simulation.  This contains four tastes for each flavour.
This enables us to perform simulations with $N_f \geq 12$ in this SU(3) gauge theory with twisted boundary condition.

\section{Quenched QCD case}\label{sec:quenched}
Before carrying out the simulation for $N_f=12$, we first measure the TPL running coupling in quenched QCD.
The gauge configurations are generated by the pseudo-heatbath algorithm
and overrelaxation algorithm mixed in the ratio 1:5.  One such a combination is called 
a "sweep" in the following.
In order to generate the configurations with the twisted boundary condition
we use the trick~\cite{LW:1986} proposed by L\"uscher and Weisz.
To reduce large statistical fluctuation of the TPL coupling, as reported
in Ref.~\cite{Divitiis:1995}, 
we measure Polyakov loops at every Monte Calro sweep and perform 
a jackknife analysis with large bin size, typically of $O(10^3)$.
This enables us to evaluate the statistical error correctly.
The simulations are carried out with several lattice sizes ($L/a=4,6,8,10,12,14,16$)
at more than twenty $\beta$ values in the range $6.2 \leq \beta \leq 16$.
We generate 200,000-400,000 configurations for each $(\beta, L/a)$ combination.
\begin{figure}[h]
\begin{center}
  \includegraphics[height=7cm]{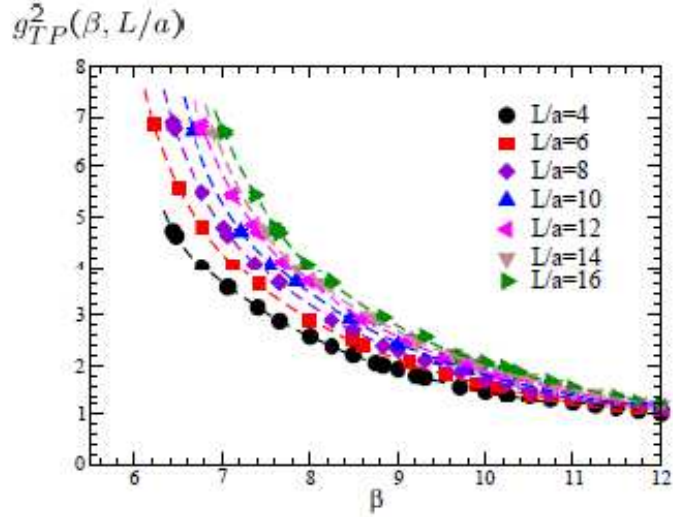} 
\end{center}
\caption{TPL renormalized coupling in the each $\beta$ and $L/a$ in quenched QCD.}
\label{fig:quenched-global-fit}
\end{figure}
Figure~\ref{fig:quenched-global-fit} shows the $\beta$ dependence of the 
renormalized coupling in TPL scheme at various lattice sizes.
The results are fitted at each fixed lattice size 
to the interpolating function which is similar to the one used in Ref.~\cite{Appelquist:2009ty},
\begin{equation}
g^2_{\mathrm{TP}}(\beta) = \sum_{i=1}^n \frac{A_i}{(\beta - B)^i},
\end{equation}
where $A_i$ are the fit parameters, and $4\leq B \leq 5$, $n=3,4$ are employed.
As a small lattice size of the step scaling, we use $L/a=4,6,8,10$.
The step scaling parameter is $s=1.5$, and we estimate the coupling constant 
for $L/a=9,15$ from interpolations at the fixed $\beta$ using the above
fit results of all the lattice sizes.

\begin{figure}[h]
\begin{center}
  \includegraphics[height=7cm]{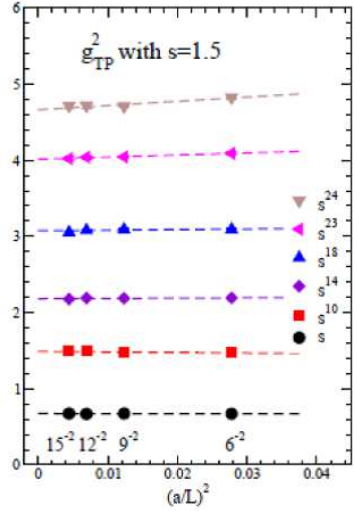} 
\end{center}
\caption{The continuum limit of $g^2_{TP}$ with $s=1.5$. The fit function is a linear 
function of $(a/L)^2$. The statistical error bars are of the same size of the symbols.}
\label{fig:quenched-cont-lim}
\end{figure}

We take the continuum limit using a linear function in $(a/L)^2$, because the 
TPL scheme ivolves no $O(a/L)$ error.
We found that the coupling constant of the TPL scheme exhibits scaling 
behaviour
even at the small lattice sizes, as shown in Fig.\ref{fig:quenched-cont-lim}.

\begin{figure}[h]
\begin{center}
  \includegraphics[height=8cm]{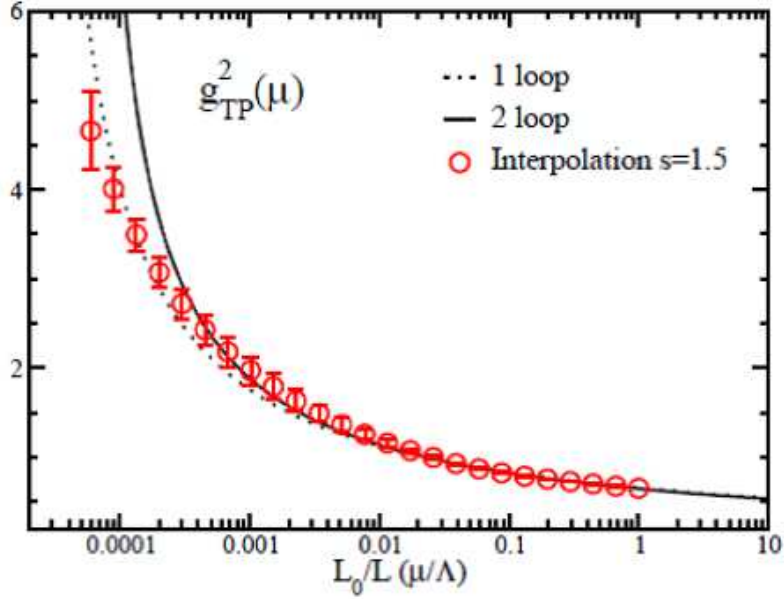} 
\end{center}
\caption{The running coupling constants in TPL scheme, one-loop and two-loop.}
\label{fig:quenched-running}
\end{figure}

The TPL running coupling constant in quenched QCD with 24 steps
is shown in Fig.\ref{fig:quenched-running} together with one- and two-loop perturbative results.
The horizontal axis corresponds to the energy scale.
All the results are normalized at $L=L_0$ with $g^2(L_0/L) = 0.65$.
The nonperturbative running coupling constant is consistent with one- and two-loop perturbative 
results in the high energy region ($L_0/L \geq 0.1$).
On the other hand, in the low energy region, the running is slower than one-loop. This shows 
the feature of TPL scheme.
The TPL running coupling constant in $\mu=1/L \rightarrow 0$ limit goes to $1/k \sim 32$, since 
the boundary effects becomes negligible in this limit. Thus the definition of eq.(\ref{TPL-def}) 
goes to the constant.
This is the reason why the nonperturbative running coupling constant in this scheme runs slower than the 
one-loop perturbative result in the low energy region.
From this quenched test, we conclude that we can control both the
the statistical and systematic errors of the TPL coupling constant,
and can obtain reasonable result with this scheme.
Furthermore we found the TPL coupling constant in quenched QCD
has a robust scaling behaviour even in a small lattice size, which was also
observed in the previous quenched SU(2) calculations~\cite{deDivitiis:1993hj,Divitiis:1995}.

\section{$N_f=12$ case}\label{sec:Nf-12}
In this section, we present preliminary results for our nonperturbative running coupling constants. 
It is consistent with the pertubartive result at high energy.

\begin{figure}[h]
\begin{center}
  \includegraphics[height=9cm]{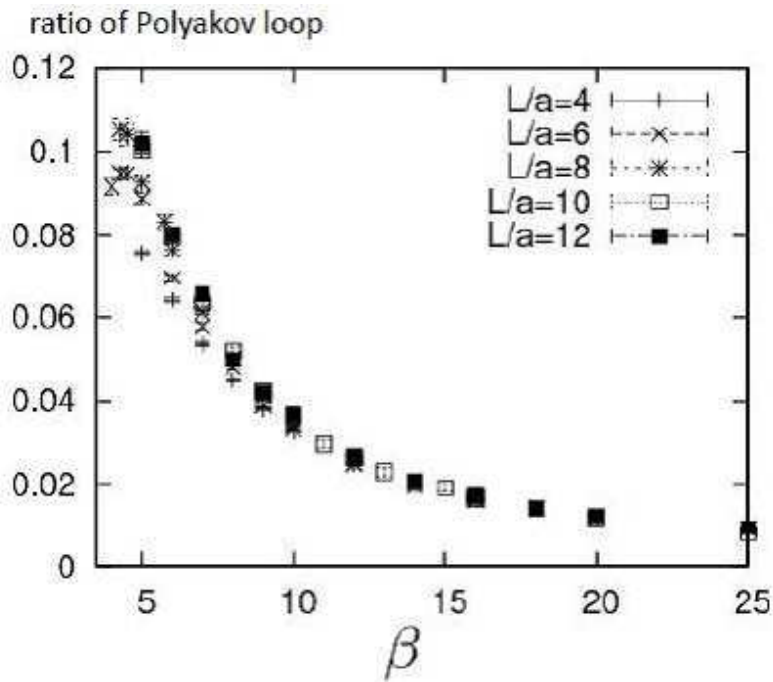} 
\end{center}
\vspace{-0.6cm}
\caption{The ratio of Polyakov loop in each $\beta$ and $L/a$. }
\label{fig:Nf-12-global}
\end{figure}

The simulation parameters are $4.0 \leq \beta \leq 25.0$ with lattice sizes $L/a=4,6,8,10,12$ 
\footnote{At the lattice conference, we took the UV starting point of step scaling at $g^2=0.542$,
a large value.  Therefore the running behavior was not consistent with the perturbative results even at this UV starting point.
In this proceedings, we will report the modified result which has been obtained using a starting point much deeper into the UV regime.}.
Figure~\ref{fig:Nf-12-global} shows the global behavior of the ratio of Polyakov loops 
in eq.(\ref{TPL-def}) for each lattice size.  This is also the global behaviour of the TPL coupling constant.
Note that the behavior in low-$\beta$ region is different from the SF scheme \cite{Appelquist:2009ty}, and the TPL 
coupling shows the trend of reaching a plateau for each lattice size.
The effect of taste breaking for staggered fermions results in significant scheme dependence in this region.

It is difficult to find a good interpolating function to fit all the data.  This is due to the plateau behaviour
in the IR.
In this proceedings, we use the 38 data points in the high${-}\beta$ regime, where the ratios of 
Polyakov loops are smaller than $0.04$, to perform the global fit to the interpolating function $f(x,y)=f(\beta,a/L)$:
\beq
f(x,y)=\frac{6k}{x+c_1 log(y)}+\frac{c_2+c_3 log(y)}{(x+c_1 log(y))^2}.
\eeq  
Here, we fix the coefficient of the first term to be $6k$. This is because the renormalised 
coupling constant should be equal to the bare one in the UV (high $\beta$).
From the perturbative analysis at high energy, we can fix $a/L$ dependence.

\begin{figure}[h]
\begin{center}
  \includegraphics[height=8cm]{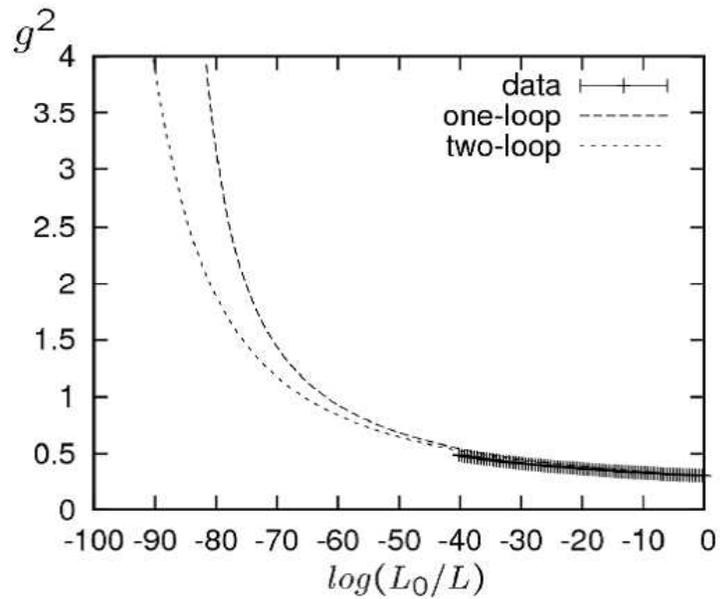} 
\end{center}
\caption{The TPL running coupling constant in high energy region.}
\label{fig:Nf-12-running}
\end{figure}

We carry out step scaling procedure similar to that in the quenched case.
Figure~\ref{fig:Nf-12-running} shows the running of our $99$ steps starting at the UV point $g^2=0.298$, 
which corrresponds to the ratio of Polyakov loop being $0.009487$.  Perturbative results are also shown
in this plot.
We find good agreements with perturbative running in this UV regime.
Presently, we are investigating the issue of finding a good interpolating function to describe the $\beta$-dependence 
in the IR.  This is important for a detailed study of the IR fixed point.
We find that as a function of $\beta$ the value of the running coupling for
fixed $L/a=4,6$ stops growing towards smaller $\beta$ at around $\beta=4.5$
and deviate from the larger volume data. This is in contrast to the case
of SF scheme where the running couplings for each L/a continue to grow
towards smaller $\beta$ and cross with each other around $\beta=4.5$.

The step scaling analysis in the strong coupling region is in progess.

\section*{Acknowledgements}
Numerical simulation was carried out on the vector supercomputer
NEC SX-8 at YITP, Kyoto University, and SX-8 at RCNP, Osaka University.
This work is supported in part of the Grant-in-Aid of the Ministry of Education (Nos. 19540286, 19740121, 19740160, 19GS0219, 20105002, 20740133, 20105005, 21105508 and 21-897).
E.B. is supported by the EU-project 227431: FP7 HADRON PHYSICS-II.
M.K. is supported by the U.S. Department of Energy at Los Alamos National Laboratory
under Contract No. DE-AC52-06NA25396.
C.-J.D.~L. is supported by the National Science Council of Taiwan via grant 96-2122-M-009-020-MY3.

\end{document}